# Image Fragile Watermarking Algorithm Based on Deneighborhood Mapping

Yilong Wang[1], Zhenyu Li[1], Daofu Gong[1,*], Haoyu Lu[1], Fenlin Liu[1]
[1]The Province Key Laboratory of Cyberspace Situation Awareness,
Zhengzhou 450001, China
*E-mail: gongdf@aliyun.com

Abstract: To address the security risk caused by fixed offset mapping and the limited recoverability of random mapping used in image watermarking, we propose an image self-embedding fragile watermarking algorithm based on deneighborhood mapping. First, the image is divided into several $2\times 2$ blocks, and authentication watermark and recovery watermark are generated based on the average value of the image blocks. Then, the deneighborhood mapping is implemented as, for each image block, its mapping block is randomly selected outside it's neighborhood whose size is specified by a parameter. Finally, the authentication watermark and the recovery watermark are embedded in the image block itself and its corresponding mapping block. Theoretical analysis indicates that in the case of continuous region tampering, the proposed watermarking method can achieve better the recovery rate of the tampered image block than the method based on the random mapping. The experimental results verify the rationality and effectiveness of the theoretical analysis. Moreover, compared with the existing embedding algorithms based on random mapping, chaos mapping and Arnold mapping, in the case of continuous region tampering, the average recovery rate of the tampered region achieved by the proposed algorithm is higher.

Keywords：fragile watermark, self-embedding, deneighborhood mapping

**Supported by**: National Natural Science Foundation of China (62002387, u1736214, 61872448, 61772549)

# 1. Introduction

As the main carrier for acquiring and disseminating information, digital images bring great convenience to people's lives, but their easy editing and modification characteristics also bring a series of security risks. How to confirm the authenticity of digital images, locate, and restore the tampered area has become an important research content in the information security community. Self-embedding watermarking technology encodes the image information as watermark information and embeds it in the image. It checks the change of watermark in the tampered image, authenticates the integrity, and recovers the tampered image area through the encoded information of the image itself.

In the research of self-embedding watermarking technology, block-based watermark generation and embedding is commonly used. By selecting the mapping sub block of the image block and embedding the restoration watermark of the image block in its mapping image block, the restoration of the tampered image block is realized. The selection methods of mapping image blocks can be divided into two categories: One is the translation like mapping method. For example, Kim *et al.* [1] proposed a self-embedding fragile watermark detection scheme based on absolute block truncation coding and optimal pixel adjustment to detect image tampering, and selected the mapping block by means of a fixed offset. Singh *et al.* [2] proposed an effective self-embedding watermarking scheme for image tampering detection and location with recovery capability, and found the mapping block by means of a fixed offset. This algorithm has two levels of authentication, and the authentication result is better. Feng *et al.* [3] proposed a semi fragile digital watermarking algorithm for interference image authentication and recovery. The algorithm divides the image into $2 \times 2$ blocks and uses a fixed offset to find the mapping block. Rakhmawati *et al.* [4], Wang *et al.* [5], and Singh *et al.* [6] all used the same embedding method, whereas the embedding method of Shi *et al.* [7] fixed the offset of the image block in the upper half of the image to the lower half of the image and fixedly shifted the lower half of the image block to the upper half. This kind of fixed offset image block mapping algorithm sacrifices the randomness of watermark embedding, which may lead to security risks of the algorithm.

The other category is random mapping. Lee *et al.* [8] proposed a self-embedded authentication watermark high-quality image recovery algorithm, which generates a random sequence through a key and finds the mapping block according to the random sequence. The embedding method of Dadkhah *et al.* [9] randomly maps the image blocks in the upper half of the image to the lower half of the image and randomly maps the image blocks in the lower half to the upper half. Zhang *et al.* [10] proposed a new image tampering location and recovery algorithm based on watermarking technology. The algorithm image is divided into $2 \times 2$ image blocks, and the chaotic map is used to select the mapping block. This algorithm uses the average value and feature of the block to generate two authentication watermarks. Tong *et al.* [11], Hemida *et al.* [12] and Dhole *et al.* [13] also selected the mapping block based on chaos mapping. Al-Out *et al.* [14] proposed a tamper detection and content restoration image watermarking scheme based on scrambling–preserving two-level moment method. According to Arnold scrambling, it selects the mapping block, and the image restoration effect of this algorithm is more advantageous. The recovery rate of this kind of random image block mapping algorithm is limited when the image area is tampered with.

In summary, based on the block-based self-embedding watermark algorithm, when selecting the watermark embedding mapping block, two types of approaches can be used, namely, random mapping and translation mapping. On the one hand, the random mapping method can ensure that it is difficult for the attacker to establish the corresponding relationship between image blocks without the key, and the algorithm has high security. However, this embedding method makes the image block and its mapping block randomly tampered at the same time. Therefore, when the image is tampered with continuous regions, the recoverability of the image block is equivalent to that when the image is randomly tampered. That is to say, the random selection of the mapping method can ensure the security of the algorithm, but the recoverability is limited. On the other hand, the translation mapping strategy makes the image block have a certain distance from its mapping block by setting a certain offset. Thus, the possibility that the image block and its mapping block are tampered at the same time is reduced, and the recoverability of the image block is improved. However, this method destroys the randomness of the selection of the mapping blocks and enables an attacker to obtain the correspondence between the image blocks and implement a targeted attack to destroy the watermark.

Therefore, when selecting the mapping block, a certain distance between the image block and its mapping block should be kept to prevent it from being modified at the same time. A certain randomness is also needed in the selection of mapping block to ensure the security of the algorithm. Image tampering has a certain purpose, and it is often regional tampering. To this end, this paper proposes an image self-embedding fragile watermarking algorithm based on deneighborhood mapping. First, a block neighborhood is set with the image block as the center, and then the mapping block is randomly selected from the outside of the neighborhood as the embedding position for the restoring watermark. Thus, image recoverability and algorithm security can be guaranteed at the same time. In the cases of different sizes of neighborhood and tampered area, this paper analyzes the recoverability of image blocks and verified the rationality and the validity of the proposed method.

## 2. The proposed watermark algorithm

This paper proposed a self-embedding watermarking algorithm for images without neighborhood mapping, which mainly includes two parts: watermark generation and embedding, tampering authentication and restoration. Watermark generation and embedding includes image blocking, mapping block selection, and watermark embedding. Tamper detection and recovery includes image blocking, watermark extraction, watermark generation, tamper authentication, and tamper recovery. The algorithm is specifically described as follows:

### 2.1 Generation and embedding of watermark

Assuming that the original image is $I$ and the size is $N \times N$, the watermark generation and embedding process is shown in Fig. 1.

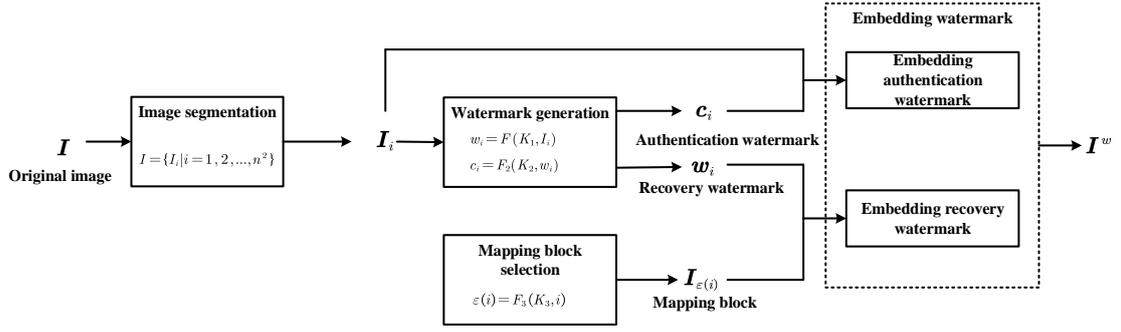

Fig. 1. Generation and embedding of watermark

1. Image blocking

The original image without overlap $I$ is divided into $n \times n$ nonoverlapping $2 \times 2$ image blocks, expressed as

$$I = \{I_i | i = 1, 2, ..., n^2\} \quad (1)$$

where $I_i$ is the block of $2 \times 2$, and $n = \dfrac{N}{2}$.

2. Watermark generation

For any image block $I_i$, the average value of all its pixels at 6-bit height is calculated, and restored watermark $w_i$ is generated, as shown in Equation (2):

$$w_i = F(K_1, I_i) \quad (2)$$

Hash treatment is carried out on $w_i$ using key $K_2$ to generate a 2-bit authentication watermark $c_i$.

$$c_i = F_2(K_2, w_i) \quad (3)$$

3. Mapping block selection

Image block $I_i$ is taken as the center to set a neighborhood $R_{I_i}^r$ with a size of $r \times r$ ($r$ is an odd number), as shown in Fig. 2.

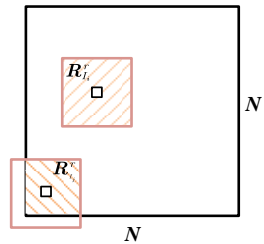

Fig. 2. Schematic diagram of the neighborhood

where the shaded part represents $R_{I_i}^r$. According to key $K_3$, an image block $I_{\varepsilon(i)}$ randomly selects from the area outside neighborhood $R_{I_i}^r$ as the mapping block of $I_i$, $\varepsilon(i) = F_3(K_3, i)$, and $I_{\varepsilon(i)}$ satisfies the following: For any $i \neq j$, $\varepsilon(i) \neq \varepsilon(j)$.

4. Watermark embedding

The authentication watermark $c_i$ of image block $I_i$ in its lowest 2-bit is embedded, and restoration watermark $w_i$ is embedded in the lowest 2-bit of its mapping block $I_{\varepsilon(i)}$ to generate a watermarked image $I^w$.

## 2.2 Authentication and recovery

Assuming that the watermarked image to be detected is $I'$, tampering authentication and recovery is shown in Figure 3.

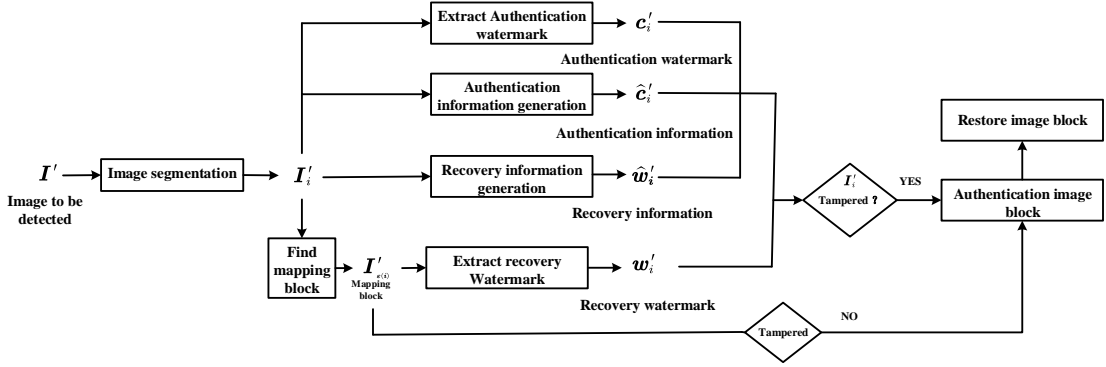

Fig. 3. Schematic diagram of tampering authentication and recovery

1. Image blocking

$I'$ nonoverlapping image blocks are divided into image blocks of size $2 \times 2$, expressed as

$$I' = \{I'_i | i = 1, 2, ..., n^2\}. \tag{4}$$

2. Watermark extraction

For image block $I'_i$, authentication watermark $c'_i$ is extracted from the lowest bit of its pixel. Using the mapping method in 2.1, mapping block $I'_{\varepsilon(i)}$ is found to extract authentication watermark $c'_{\varepsilon(i)}$ of mapping block $I'_{\varepsilon(i)}$ and restoration watermark $w'_i$ of image block $I'_i$ from the lowest bit of the pixel.

3. Watermark generation

For image block $I'_i$, the authentication watermark generation method in 2.1 is used to generate its recovery information $\hat{w}'_i$ and authentication information $\hat{c}'_i$. The authentication information

$\widehat{c}'_{\varepsilon(i)}$ of image block $I'_{\varepsilon(i)}$ is generated.

4. Authentication

(1) If $c'_i \neq \widehat{c}'_i$, image block $I'_i$ has been tampered with and does not pass the authentication.

(2) If $c'_i = \widehat{c}'_i$ and $w'_i = \widehat{w}'_i$, image block $I'_i$ has not been tampered with and passes the authentication.

(3) When $c'_i = \widehat{c}'_i$ and $w'_i \neq \widehat{w}'_i$, if $c'_{\varepsilon(i)} \neq \widehat{c}'_{\varepsilon(i)}$, mapping block $I'_{\varepsilon(i)}$ of image block $I'_i$ may be tampered; thus, $I'_i$ has not been tampered, and the authentication is passed.

(4) When $c'_i = \widehat{c}'_i$ and $w'_i \neq \widehat{w}'_i$, if $c'_{\varepsilon(i)} = \widehat{c}'_{\varepsilon(i)}$ and $w'_{\varepsilon(i)} = \widehat{w}'_{\varepsilon(i)}$, the mapping block $I'_{\varepsilon(i)}$ of image block $I'_i$ has not been tampered; thus, $I'_i$ has not been tampered, and the authentication is not passed.

(5) Otherwise, when $c'_i = \widehat{c}'_i$ and $w'_i \neq \widehat{w}'_i$, if $c'_{\varepsilon(i)} = \widehat{c}'_{\varepsilon(i)}$ and $w'_{\varepsilon(i)} \neq \widehat{w}'_{\varepsilon(i)}$, mapping block $I'_{\varepsilon(i)}$ of image block $I'_i$ may be tampered; thus, $I'_i$ has not been tampered, and the authentication is passed.

(6) After all image blocks are authenticated, for any image block $I'_i$, whether an unauthenticated block is in its adjacent image block is judged. If an unauthenticated block exists, image block $I'_i$ is not authenticated; otherwise, image block $I'_i$ is considered authenticated.

5. Tamper recovery

For tampered image block $I'_i$, if its mapping image block $I'_{\varepsilon(i)}$ has not been tampered, image block $I'_i$ can be restored using Equation (5):

$$I'_i = F_1^{-1}(K_1, w'_i) \tag{5}$$

## 3. Theoretical analysis of tampering recovery rate

To evaluate the performance of this algorithm, this section provides the recovery rate of the tampered image block and the average recovery rate of the tampered area in the case of continuous area tampering from the perspective of theory and numerical calculation. For better description, the following symbols are introduced:

$L = \{B_i | i = 1, 2, \cdots, |L|\}$ indicates the tampere area of the image to be detected, where $B_i$ is the *i*-th image block of $2 \times 2$, and $|*|$ represents the collective potential operation.

$H^r(*)$ is the recovery rate of the tampered image block*.

$\bar{H}^r$ represents the average recovery rate of the tampered area $L$.

## 3.1 The theoretical recovery performance of the proposed algorithm

According to the algorithm proposed in Section 2, when the mapping block of image block $B_i$ is tampered, the image block can be restored by extracting the restoration watermark. The mapping block of image block $B_i$ is selected from the area outside its neighborhood, and the neighborhood is denoted as $R^r_{B_i}$. Only when the mapping block is also in the tampered area $L$, the image block $B_i$ cannot be restored. Therefore, the recovery rate of image block $B_i$ is

$$H^r(B_i) = 1 - \frac{|L - R^r_{B_i}|}{|I - R^r_{B_i}|} = 1 - \frac{|L| - |L \cap R^r_{B_i}|}{n^2 - |R^r_{B_i}|} \tag{6}$$

where $|L - R^r_{B_i}|$ indicates the number of tampered image blocks except for field $R^r_{B_i}$, and $|I - R^r_{B_i}|$ indicates the possible mapping area size of image block $B_i$. The average recovery rate of tampered area $L$ is

$$\bar{H}^r = 1 - \frac{1}{|L|} \sum_{i=1}^{|L|} \frac{|L| - |L \cap R^r_{B_i}|}{n^2 - |R^r_{B_i}|} \tag{7}$$

If the tampered image blocks are randomly distributed in the whole image, which is called random tampering mode, the number of tampered blocks in the neighborhood of each image block neighborhood is $|R^r_{B_i}||L|/n^2$ in theory, namely,

$$|L \cap R^r_{B_i}| = |R^r_{B_i}||L|/n^2 \tag{8}$$

Equation (6) and Equation (7) can be rewritten as

$$H^r(B_i) = 1 - \frac{|L|}{n^2}, \quad \bar{H}^r = 1 - \frac{|L|}{n^2} \tag{9}$$

According to Equation (9), the recovery performance of the proposed algorithm in the random tampering mode is equivalent to that of the algorithms based on random mapping embedding algorithm.

Considering that the purpose of image tampering is often continuous tampering, setting the tampered area to be an area with a size of $l \times l$ is better, then tampered area $L$ can be rewritten as $L = \{B_{i,j} | i, j = 1, 2, \cdots, l\}$, and $B_{i,j}$ is the image block of $2 \times 2$; image $I$ is rewritten as $I = \{I_{\alpha,\beta} | \alpha, \beta = 1, 2, \cdots, n\}$, and $I_{\alpha,\beta}$ is the image block of $2 \times 2$. The relationship between tampered area $L$ and neighborhood $R^r_{B_{i,j}}$ of image block $B_{i,j}$ is shown in Figure 4–6.

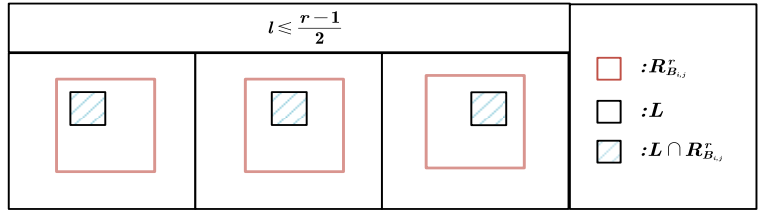

Fig. 4. Relationship between tampered area $L$ and neighborhood $R^r_{B_{i,j}}$ of image block $B_{i,j}$ at $l \leqslant \frac{r-1}{2}$

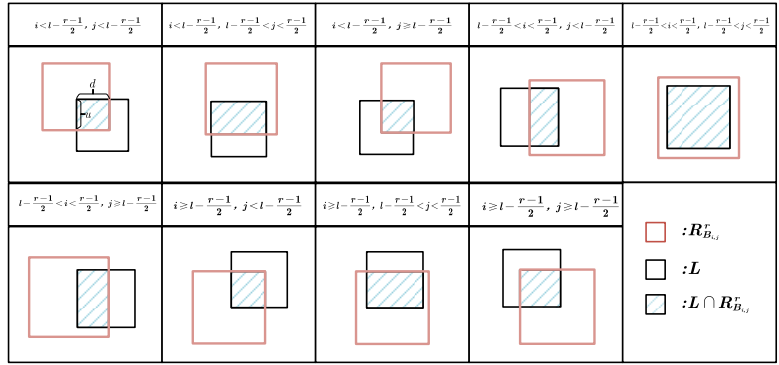

Fig. 5. Relationship between tampered area $L$ and neighborhood $R^r_{B_{i,j}}$ of image block $B_{i,j}$ at $\frac{r-1}{2} < l \leqslant r$

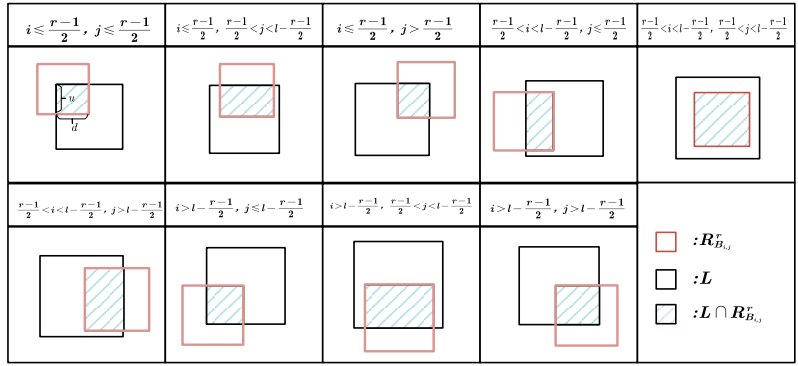

Fig. 6. Relationship between tampered area $L$ and neighborhood $R^r_{B_{i,j}}$ of image block $B_{i,j}$ at $l > r$

According to Figure 4, when $l \leqslant \frac{r-1}{2}$, for $B_{i,j} \in L$, $L \subseteq R^r_{B_{i,j}}$ is established, at this time $|L \cap R^r_{B_{i,j}}| = l^2$, then,

$$H^r(B_{i,j}) = 1, \bar{H}^r = 1 \qquad (10)$$

Figure 5 shows that when $\frac{r-1}{2} < l \leqslant r$, $|L \cap R^r_{B_{i,j}}| = u \times d$,

$$H^r(B_{i,j}) = 1 - \frac{l^2 - (u \times d)}{n^2 - |R^r_{B_{i,j}}|}, \bar{H}^r = 1 - \frac{1}{l^2} \sum_{i=1}^{l} \sum_{j=1}^{l} \frac{l^2 - (u \times d)}{n^2 - |R^r_{B_{i,j}}|} \quad (11)$$

where

$$u = \begin{cases} \frac{r-1}{2} + i, & i < l - \frac{r-1}{2} \\ l, & l - \frac{r-1}{2} < i < \frac{r-1}{2} \\ \frac{r-1}{2} + (l-i), & i \geqslant \frac{r-1}{2} \end{cases}, d = \begin{cases} \frac{r-1}{2} + j, & j < l - \frac{r-1}{2} \\ l, & l - \frac{r-1}{2} < j < \frac{r-1}{2} \\ \frac{r-1}{2} + (l-j), & j \geqslant \frac{r-1}{2} \end{cases} \quad (12)$$

Similarly, Figure 6 shows that when $l > r$, the recovery rate of tampered image block $B_{i,j}$ and the average recovery rate of tampered area $L$ are calculated as Equation (11), where the values of $u$ and $d$ are as follows: (13)

$$u = \begin{cases} \frac{r-1}{2} + i, & i \leqslant \frac{r-1}{2} \\ r, & \frac{r-1}{2} < i < l - \frac{r-1}{2} \\ \frac{r-1}{2} + (l-i), & i > l - \frac{r-1}{2} \end{cases}, d = \begin{cases} \frac{r-1}{2} + j, & j \leqslant \frac{r-1}{2} \\ r, & \frac{r-1}{2} < j < l - \frac{r-1}{2} \\ \frac{r-1}{2} + (l-j), & j > l - \frac{r-1}{2} \end{cases} \quad (13)$$

Does an appropriate block neighborhood parameter $r$ exist, so that the image block recovery rate of this algorithm is better than the random embedding algorithm in the case of continuous region tampering? This question is to examine if the following equation holds.

$$Q = H^r(B_{i,j}) - \left(1 - \frac{l^2}{n^2}\right) = \frac{n^2(u \times d) - l^2 |R^r_{B_{i,j}}|}{n^2(n^2 - |R^r_{B_{i,j}}|)} \geq 0 \quad (14)$$

where $r^2/4 \leq u \times d \leq l^2$, and for Equation (14)

$$Q > \frac{n^2(u \times d) - r^2 l^2}{(n^2 - |R^r_{B_{i,j}}|)n^2} \quad (15)$$

According to Equation (15), when $r \leq n\sqrt{(u \times d)}/l$, $Q > 0$. That means a block neighborhood exists, so that the image block recovery rate of the proposed algorithm based on de neighborhood mapping is better than the one based on random mapping in the case of continuous region tampering.

## 3.2 The influence of the position of the tampered area

According to the proposed algorithm and the image block neighborhood $R^r_{B_{i,j}}$, the following propositions are established.

Proposition 1. For two consecutive tampered areas $L_1 = \{B_{i,j} | i,j = 1, 2, \cdots, l\}$ and $L_2 = \{\tilde{B}_{i,j} | i,j = 1, 2, \cdots, l\}$ of the same size in the image, as shown in Figure 7,

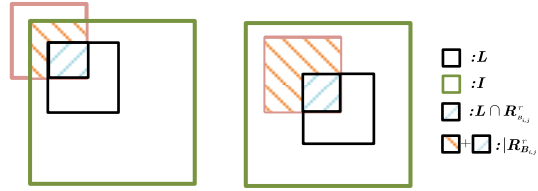

Fig. 7. Schematic diagram of different positions of *L* in the image

then the tampered area which is closer to the center of the image has higher recovery rate of each image block, namely

$$H^r(B_{i,j}) > H^r(\tilde{B}_{i,j}) \tag{16}$$

Proof: Figure 7 shows that

$$|R^r_{B_{i,j}}| < r^2 = |R^r_{\tilde{B}_{i,j}}| \tag{17}$$

Equation (11) shows that the proposition is true.

Proposition 1 shows that the recovery rate of image blocks of consecutive tampered areas of the same size has a certain correlation with the position of the tampered area in the image. That when the tampered area is the same, the situation can be inferred from Proposition 1:

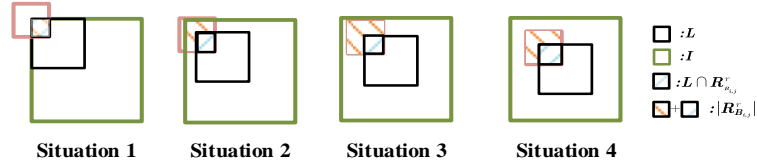

Situation 1    Situation 2    Situation 3    Situation 4

Fig. 8. Schematic diagram of three different positions of *L* in the image

where Case 1: $B_{1,1} = I_{1,1}$ (denoted as a corner); Case 2: $B_{1,1} = I_{i,j}$, $|R^r_{B_{i,j}}| < r^2$; Case 3: $B_{1,1} = I_{i,j}$, $|R^r_{B_{i,j}}| = r^2$ for any $B_{i,j}$; Case 4: the center of the tampered area *L* coincides with the center of the image *I* (denoted as the center). The recovery rate of the image block in the relative position of the tampered area and the average recovery rate of the area should be the highest in Case 1, followed by Case 2, and the lowest in Cases 3 and 4.

When the tampering parameter changes, analyzing the recovery rate of the image block as a whole is difficult. The following analysis of the tampered area from the perspective of numerical calculation is shown in Cases 1, 2, and 4. When the tampering parameter changes, the effect of the location of the tampering area on the average recovery rate is analyzed.

Figure 9 shows the numerical calculation results of the average recovery rate of the tampered area $L$ under the conditions of $B_{1,1} = I_{1,1}$, $B_{1,1} = I_{4,6}$ and the center when the image size is $512 \times 512$ and the block neighborhood parameters are $r = 101$ and $r = 201$.

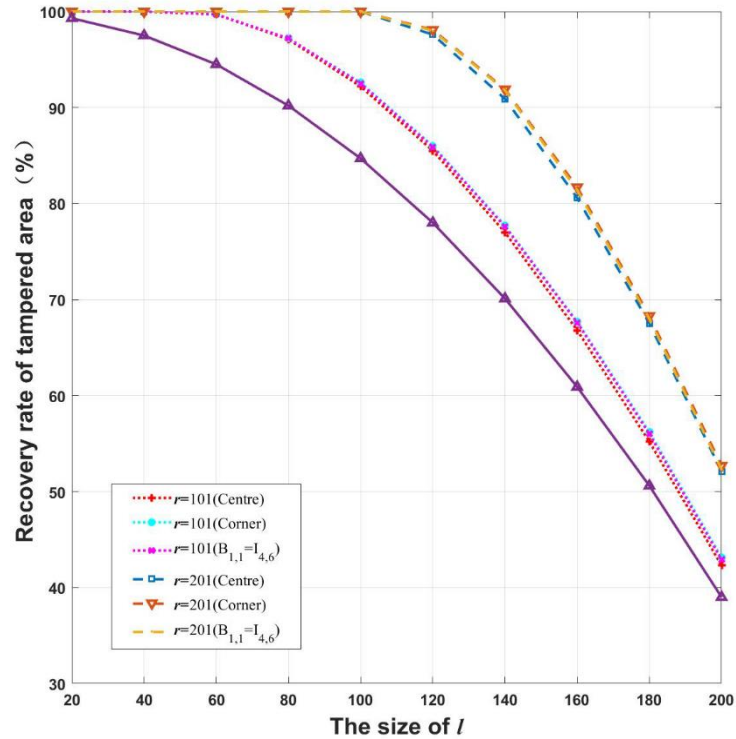

Fig. 9. Average recovery rate when the location of the tampered area $L$ is different

Figure 9 shows that the average recovery rate of tampered area $L$ has a certain correlation with the position of $L$ in the image. However, when parameter $r$ is fixed, the average recovery rate of the tampered area is very close. Therefore, when the tampering parameter $l$ is the same, the influence of tampered area's position on the performance of the algorithm can be ignored.

## 3.3 The influence of the neighborhood parameters

Next, when the values of block neighborhood parameter r and tamper parameter $l$ are different, the change trend of average recovery rate of tamper area is analyzed from the perspective of numerical calculation.

Figure 10 shows the numerical calculation results of the average recovery rate of the tampered area when the image size is $512 \times 512$ and block neighborhood parameter $r$ in tampered area $L$ and tampering parameter $l$ are different $B_{1,1} = I_{4,6}$.

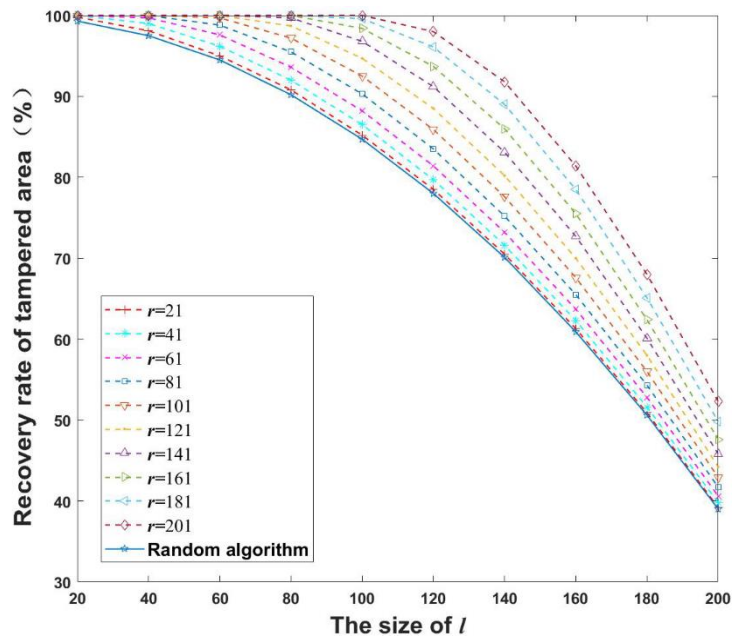

Fig. 10. Average recovery rate of tampered area at different values of parameters $r$ and $l$

Figure 10 shows that when tampering parameter $l$ is fixed, the average recovery rate of the tampered area increases as block neighborhood parameter $r$ increases; when block neighborhood parameter $r$ is fixed, the average recovery rate of the tampered area shows a downward trend as tampering parameter $l$ increases. However, a suitable parameter $r$ must exist, and thus, the average recovery rate of the proposed algorithm is higher than that of random mapping.

Figure 11 shows the numerical calculation results of recovery rate $H^r(B_{i,j})$ of block image $B_{i,j}$ when image size is $512 \times 512$, the block neighborhood parameters $r$ is set to 31 and 101 in tampered area $L$, and tampering parameter $B_{1,1} = I_{4,6}$ is $l = 40 \sim 160$.

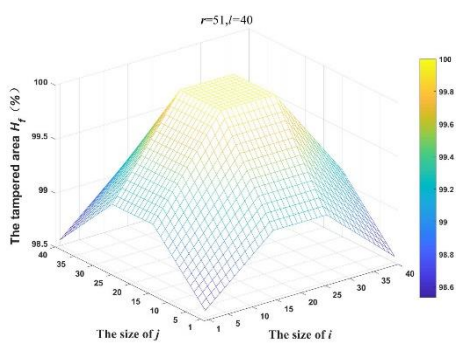 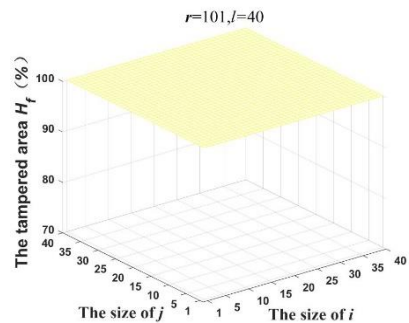

Fig.11(a) Image block recovery rate($r = 51, l = 40$)  Fig.11(b) Image block recovery rate($r = 101, l = 40$)

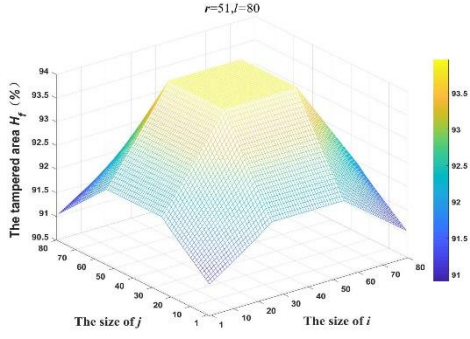 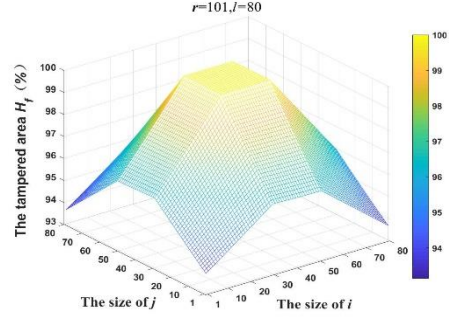

Fig.11(c) Image block recovery rate($r = 51, l = 80$)    Fig.11(d) Image block recovery rate($r = 101, l = 80$)

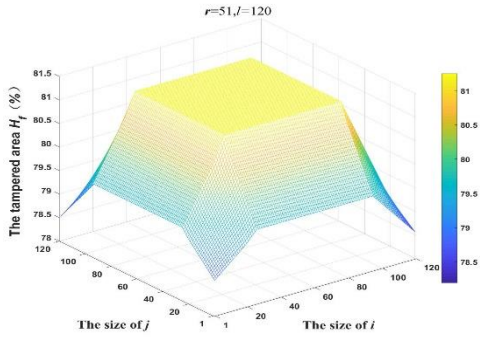 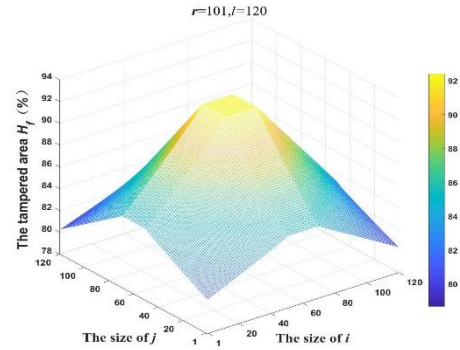

Fig.11(e) Image block recovery rate($r = 51, l = 120$)    Fig.11(f) Image block recovery rate($r = 101, l = 120$)

Fig. 11. Numerical distribution of the recovery rate of the tampered image block at $B_{1,1} = I_{4,6}$

Figure 11 shows that when the tampering parameters are fixed, the recovery rate of the tampered area image block increases with the increase of the neighborhood. When $l \leq (r-1)/2$, it can be seen from Figure 11(b) that the tampered area image block can be 100% recovered. When $(r-1)/2 < l \leq r$, it can be seen from Figure 11(a) and (d) that some image blocks in the center of the tampered area can be recovered 100%, and the recovery rates of other image blocks are more than 90%. When $l > r$, it can be seen from Figure 11(e) and (f) that the image block recovery rate in the tampered area decreases, but it is still higher than the random mapping embedding algorithm. Figure 11 also shows that the larger the intersection between the neighborhood of the image block and the tampered area, the higher the recovery rate of the image block. In another word, the closer the image block to the center of the tampered area, the higher the recovery rate, which is very beneficial to infer the tampering intention.

In summary, a suitable neighborhood exists ensuring a better recovery performance than random mapping algorithm under continuous large area tampering. For continuous tampering with a fixed size, the average recovery performance of the region increases as the neighborhood increases. The more image blocks in the neighborhood of the image block are tampered, the better the recovery performance, and the algorithm shows good adaptability to regional tampering.

# 4. Experimental results and comparison

12 images are used as the experimental object, all of which were the grayscale images of $512 \times 512$, as shown in Figure 12.

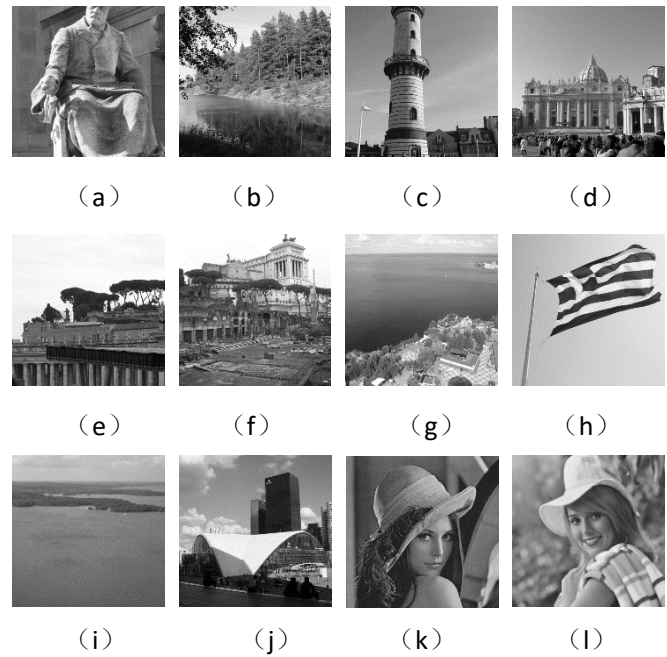

(a)　　　(b)　　　(c)　　　(d)

(e)　　　(f)　　　(g)　　　(h)

(i)　　　(j)　　　(k)　　　(l)

Fig. 12. Test images

Ten images from Figure 12 (a-j) are used to verify the experimental results and theoretical analysis results of the algorithm. Figure 12 (k) is used to test the region tamper recovery effect of our algorithm. Figure 12 (l) is used to compare the restoration effect of this algorithm with random mapping watermark embedding algorithms such as [8], [14] and [7].

## 4.1 Performance of this algorithm

For the 10 test images in Figure 12, five values for neighborhood parameters $r$ ($r = 21 \sim 101$ (step size 20)) were selected, and a different key was used for each image to generate a watermarked image. Five values for tampering parameters $l = 20 \sim 100$ (step 20) were selected for the image after the watermark was embedded. The tampering is made by cropping a square region whose size is $l \times l$. The left-up corner of the tampered region is set to $B_{1,1} = I_{4,6}$. The average recovery rate of the tampered area of 10 watermarked images under the same neighborhood parameters and the same tampering parameters was taken as the experimental results. Table 1 shows the comparison between the theoretical calculation results and the experimental results under the same location of the tampered area.

Table 1. Average recovery rate of the tampered area of the algorithm (%)

|  |  | l=20 | l=40 | l=60 | l=80 | l=100 |
|---|---|---|---|---|---|---|
| r=21 | Experimental results | 99.8 | 98.1 | 95.0 | 90.8 | 85.0 |
| | Theoretical value | 99.8 | 98.1 | 95.0 | 90.8 | 85.2 |
| r=41 | Experimental results | 100 | 99.0 | 96.2 | 92.0 | 86.3 |
| | Theoretical value | 100 | 99.0 | 96.2 | 92.0 | 86.5 |
| r=61 | Experimental results | 100 | 99.6 | 97.5 | 93.6 | 88.1 |
| | Theoretical value | 100 | 99.7 | 97.6 | 93.6 | 88.2 |
| r=81 | Experimental results | 100 | 100 | 98.8 | 95.4 | 90.1 |
| | Theoretical value | 100 | 100 | 98.8 | 95.5 | 90.3 |
| r=101 | Experimental results | 100 | 100 | 99.8 | 97.1 | 92.3 |
| | Theoretical value | 100 | 100 | 99.7 | 97.2 | 92.5 |

The table shows that under the same $l$, the average recovery rate of the tampered area increases with the increase of $r$; under the same $r$, the average recovery rate of the tampered area decreases with the increase of $l$; the maximum error between theoretical and experimental results is 0.2%. This verified the validity of the theoretical analysis.

Next, Figure 12(k) was tested to embed watermarks in the neighborhood parameters $r=101$ and $r=51$, obtaining Figure 13 (a). Then it is tampered into Figure 13 (b).

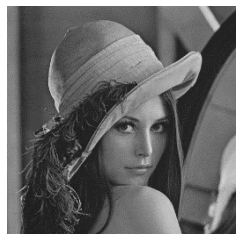 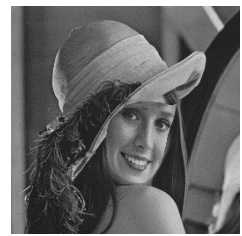

(a) Watermarked image     (b) Tampered image

Fig.13. Tampering with watermarked image

The recovery of the tamptered image achieved by the proposed algorithm, when the neighborhood parameters $r=101$ and $r=51$, are shown in Figures 14 and 15.

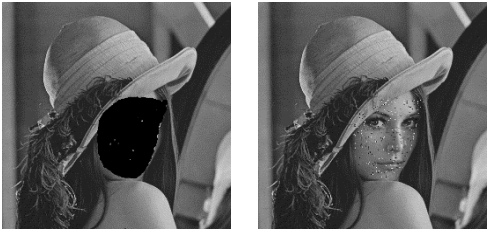 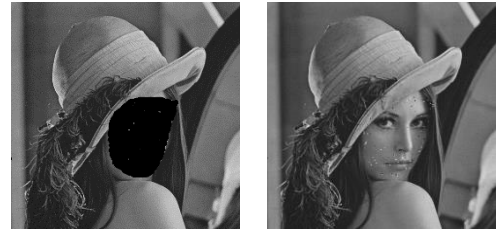

Fig.14. Authentication and recovery result($r = 51$)     Fig.15. Authentication and recovery result($r = 101$)

Figures 14 and 15 show our algorithm has a good recovery effect on continuous region tampering. When the size of the tampered area is the same, the larger the *r*, the better the recovery effect. Meanwhile, the image blocks in the center of the tampered area is more likely to be recovered than the blocks in the edge of the tampered area. This is consistent with the theoretical analysis and numerical calculation that the larger the intersection of the image block neighborhood and the tampered area, the higher the recovery rate of the image block.

## 4.2 Comparative experiment with other algorithms

The restoration effect of the proposed algorithm is compared with three other watermarking algorithm based on various mapping methods in [8], [14] and [7] for the test image from Figure 12(l). The parameter used in watermark embedding is set as $r=51$, $r=71$ and $r=101$, and watermark image (l) is tampered at various levels controlled by $l=20 \sim 100$ (the step size was 20), and the restoration result is shown in Figure 16.

It can be observed from Figure 16 that the recovery rate of the proposed algorithm in the case of continuous tampering was higher than those in literature [8], [14], and [7]; the recovery effect of the tampered area is better with the increase of $r$; and the recovery effect near the center of the tampered area is better than the edge of the area. This result further verified that the theoretical analysis is reasonable and correct.

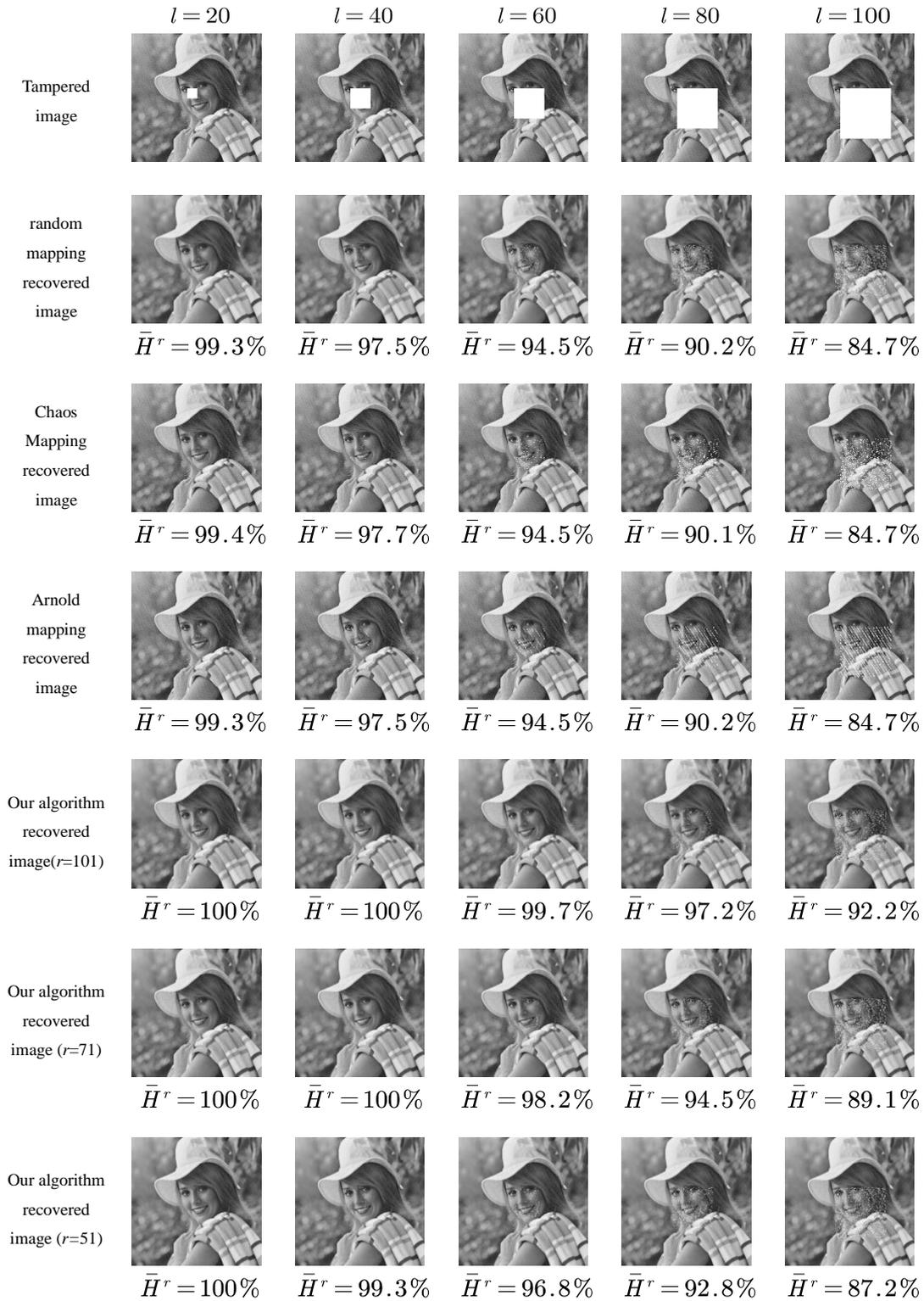

Fig. 16 Recovery comparison chart of the proposed algorithm and existing methods

The recovery performance of the proposed algorithm and algorithms in [7,8,14] are also evaluated based average recovery rate of ten images in Figure 12(a-j). The watermark embedding, tampering and recovery are applied in the same way as above. The average recovery rates of the ten images achieved by various algorithms are shown in Figure 17.

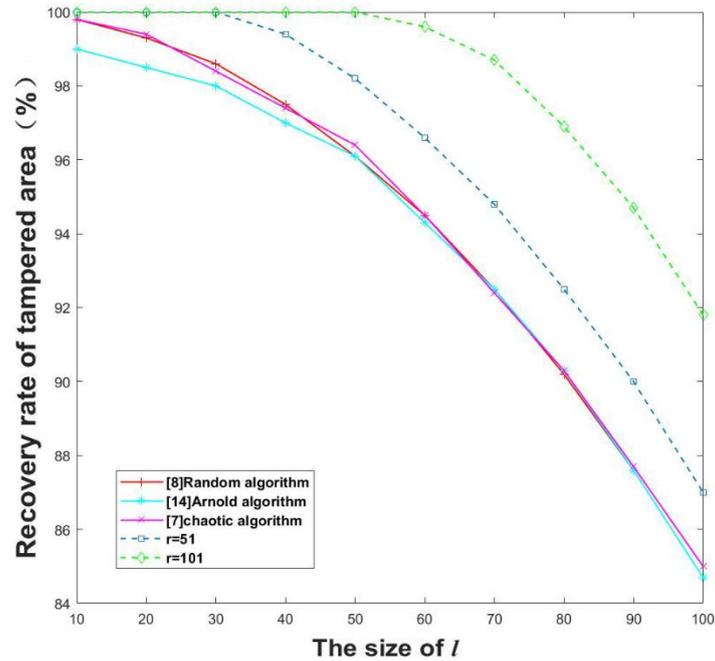

Fig.17. Comparison of tampered area recovery rates for different mapping algorithms

It can be seen from the Figure 17 that the recovery rate of the tampered area under the watermark embedding method of Arnold transform algorithm and chaotic algorithm is basically the same as that of the random map embedding watermark method. While, the proposed algorithms have higher recoverability rates when $r=51$ or $r=101$.

## 5. Conclusions

This paper proposes an image self-embedding fragile watermarking algorithm based on deneighborhood mapping. This algorithm makes the watermark embedding have a certain randomness and improves the recovery rate of image tampering. Theoretical analysis and experimental results show that in the case of continuous region tampering, when a proper neighborhood is set, the recovery rate of the tampered image block achieved by the proposed algorithm is better than that of the random mapping embedding algorithm. Besides, the recovery probability of each tampered image block increases with the intersection area between the block neighborhood and the tampered region.

## References

[1]Kim, C., Yang, C. N.: Self-Embedding Fragile Watermarking Scheme to Detect Image Tampering

Using AMBTC and OPAP Approaches. Applied Sciences.11(3), 1146(2021)

[2]Singh, D., Singh, S. K.: Effective self-embedding watermarking scheme for image tampered detection and localization with recovery capability. Journal of Visual Communication and Image Representation 38(7), 775-789(2016):

[3]Feng, B., Li X.L., Jie, Y.M., Guo, C.: A Novel Semi-fragile Digital Watermarking Scheme for Scrambled Image Authentication and Restoration. Mobile Networks and Applications. 25(12), 82–94 (2020)

[4]Rakhmawati, L., Wirawan, Suwadi. Image Fragile Watermarking with Two Authentication Components for Tamper Detection and Recovery. In : 2018 International Conference on Intelligent Autonomous Systems, pp.35-38, Singapore, 1-3 March 2018

[5]Wang, C., Zhang, H., Zhou, X.: A Self-Recovery Fragile Image Watermarking with Variable Watermark Capacity. Applied Sciences. 8(4),548(2018)

[6]Singh, D., Singh, S. K.: Block Truncation Coding based effective watermarking scheme for image authentication with recovery capability. Multimedia Tools and Applications. 78(2), 1-19(2019)

[7] Shi, H.,Wang, S.H., Li M.C., Bai, J., Feng, B.: Secure variable-capacity self-recovery watermarking scheme. Multimedia Tools & Applications. 76(5), 6941-6972(2017)

[8]Lee, C.F., Shen, J.J., Chen, Z.R.: Self-Embedding Authentication Watermarking with Effective Tampered Location Detection and High-Quality Image Recovery. Sensors (Basel, Switzerland), 19(10), 2267(2019)

[9] Dadkhah, S., Abd, M.A., Hori, Y. , Ella, H.A., Sadeghi, S.: An effective SVD-based image tampering detection and self-recovery using active watermarking. Signal Processing Image Communication 29(10), 1197-1210(2014)

[10]Zhang, J.P, Zhang, Q.F., Lv, H.L.: A novel image tamper localization and recovery algorithm based on watermarking technology. Optik, 124(23), 6367-6371(2013)

[11] Tong,X.J., Liu, Y., Zhang, M., Chen, Y.: A novel chaos-based fragile watermarking for image tampering detection and self-recovery. Signal Processing: Image Communication 28(3),301-308(2013)

[12]Hemida, O., He. H.J.: A self-recovery watermarking scheme based on block truncation coding and quantum chaos map. Multimedia Tools and Applications 79(3), 18695-18725 (2020)

[13] [1] Dhole, V.S., Patil, N.N.: Self Embedding Fragile Watermarking for Image Tampering Detection and Image Recovery Using Self Recovery Blocks. In: 2015 International Conference on

Computing Communication Control and Automation. pp. 752-757, Pune, India, 26-27 February 2015

[14] [1] Al-Otum, H.M.M., Alansari, A.W.: Image Watermarking for Tamper Detection and Content Recovery Applications Based on a Preservative Bi-Level Moment Approach with Shuffling. In: Proceedings of the 2019 5th International Conference on Computer and Technology Applications. pp. 25-29, Istanbul, Turkey, 16–17 April 2019